%% file: sdsspsf.tex
\documentclass[preprint]{aastex62}

\newcommand{\vk}{von K\'{a}rm\'{a}n}

\shorttitle{A Study of the SDSS PSF}
\shortauthors{Xin et al.}

\begin{document}

\title{A Study of the Point Spread Function in SDSS Images}

\correspondingauthor{Bo Xin}
\email{bxin@lsst.org}

\author{Bo Xin}
\affiliation{Large Synoptic Survey Telescope, Tucson, AZ 85719}

\author{\v{Z}eljko Ivezi\'{c}}
\affiliation{Department of Astronomy, University of Washington, Seattle, WA 98195}

\author{Robert H. Lupton}
\affiliation{Department of Astrophysical Sciences, Princeton University, Princeton, NJ 08544}

\author{John R. Peterson}
\affiliation{Department of Physics and Astronomy, Purdue University,
  West Lafayette, IN 47907}

\author{Peter Yoachim}
\affiliation{Department of Astronomy, University of Washington,
  Seattle, WA 98195}

\author{R. Lynne Jones}
\affiliation{Department of Astronomy, University of Washington,
  Seattle, WA 98195}

\author{Charles F. Claver}
\affiliation{Large Synoptic Survey Telescope, Tucson, AZ 85719}

\author{George Angeli}
\altaffiliation{Current address: Giant Magellan Telescope Organization, Pasadena, CA 91107}
\affiliation{Large Synoptic Survey Telescope, Tucson, AZ 85719}



\begin{abstract}
We use SDSS imaging data in $ugriz$ passbands to study the shape of the
point spread function (PSF) profile and the variation of its width with 
wavelength and time. We find that the PSF profile is well described by 
theoretical predictions based on \vk's turbulence theory. The observed
PSF radial
profile can be parametrized by only two parameters, the profile's full width
at half maximum (FWHM) and a normalization of the contribution of an empirically determined 
``instrumental'' PSF. The profile shape is very similar to the ``double gaussian
plus power-law wing'' decomposition used by SDSS image processing pipeline, 
but here it is successfully modeled with two free model parameters, rather than six as in SDSS pipeline. 
The FWHM variation with wavelength follows the
$\lambda^{\alpha}$ power law, where $\alpha \approx-0.3$ and is correlated
with the FWHM itself. The observed behavior is much better described by \vk's turbulence 
theory with the outer scale parameter in the range 5--100 m, than by the 
Kolmogorov's turbulence theory. We also measure the temporal and angular
structure functions for FWHM and compare them to simulations and
results from literature. The angular structure function saturates at scales beyond 0.5$-$1.0 degree. 
The power spectrum of the temporal behavior is found to be broadly consistent with 
a damped random walk model with characteristic timescale in the range $\sim5-30$ minutes, 
though data show a shallower high-frequency behavior. The latter is well fit 
by a single power law with index in the range $-1.5$ to $-1.0$. A hybrid model 
is likely needed to fully capture both the low-frequency and high-frequency 
behavior of the temporal variations of atmospheric seeing. 
\end{abstract}




\section{Introduction}

\input{intro} 

\section{Data Overview}

\input{data_overview}

\section{The PSF profile analysis}

\input{PSFprofile}

\section{The analysis of FWHM behavior}

\input{FWHManalysis}

\section{Discussion and Conclusions}

\input{conclusions}

\acknowledgments

This material is based upon work supported in part by the National Science Foundation through
Cooperative Agreement 1258333 managed by the Association of Universities for Research in Astronomy
(AURA), and the Department of Energy under Contract No. DE-AC02-76SF00515 with the SLAC National
Accelerator Laboratory. Additional LSST funding comes from private donations, grants to universities,
and in-kind support from LSSTC Institutional Members.

Funding for the SDSS and SDSS-II has been provided by the Alfred
P. Sloan Foundation, the Participating Institutions, the National
Science Foundation, the U.S. Department of Energy, the National
Aeronautics and Space Administration, the Japanese Monbukagakusho, the
Max Planck Society, and the Higher Education Funding Council for
England. The SDSS Web Site is http://www.sdss.org/.

\bibliographystyle{aasjournal}
\bibliography{ref}

\end{document}

%% file: intro.tex

The atmospheric seeing, the point-spread function (PSF) due to atmospheric turbulence, plays
a major role in ground-based astronomy \citep{Roddier1981}. An adequate description 
of the PSF is critical for photometry, star-galaxy separation, and for unbiased measures of 
the shapes of nonstellar objects \citep{Lupton2001}. In addition, better understanding of the 
PSF temporal variation can lead to improved seeing forecasts; for example, such forecasts are 
considered in the optimization of LSST observing strategy \citep{LSSToverview}.

Seeing varies with the wavelength of observation, and it also varies with time, on time 
scales ranging from minutes to years. These variations, as well as the radial seeing
profile, can be understood as manifestations of atmospheric instabilities due to turbulent layers. 
Although turbulence is a complex physical phenomenon, the basic properties of the atmospheric
seeing can be predicted from first principles \citep{Racine2009}. The
\vk~turbulence theory, 
an extension of the Komogorov theory that introduces a finite maximum 
size for turbulent eddies (the so-called
outer scale parameter), quantitatively predicts the seeing profile and the variation of seeing
with wavelength \citep{vk1, vk2}. Therefore, seeing measurements can be used to test the theory and estimate
the relevant physical parameters. 

An unprecedentedly large high-quality database of seeing measurements was delivered by the Sloan Digital Sky Survey (SDSS, \citealt{York2000}), a large-area multi-bandpass digital sky survey. The SDSS delivered homogeneous and deep 
($r\la22.5$) photometry in five bandpasses ($u$, $g$, $r$, $i$, and $z$, with effective wavelengths 
of 3551, 4686, 6166, 7480, and 8932 $\AA$), accurate to about 0.02 mag for unresolved sources 
not limited by photon statistics \citep{Sesar2007}. Astrometric positions are accurate to better 
than 0.1 arcsec per coordinate for sources with $r<20.5$ \citep{Pier2003}, and the morphological 
information from the images allows reliable star-galaxy separation to $r<21.5$ \citep{Lupton2002}.
 
The SDSS camera (Gunn et al. 1998) used drift-scanning observing mode (scanning along great circles
at the sidereal rate) and detected objects in the order 
$r$-$i$-$u$-$z$-$g$, with detections in two successive bands separated in time by 72 s. Each of the six camera
columns produces a 13.5 arcmin wide scan; the scans are split into ``fields'' 9.0 arcmin long, corresponding
to 36 seconds of time (the exposure time is 54.1 seconds because the sensor size is 2k by 2k pixels,
with 0.396 arcsec per pixel). 
The point-spread function (PSF) is estimated as a function of position
within each field, although we only use one estimate per field, and in each bandpass - there are about 
148 seeing estimates for each square degree of scanned sky. As a result, the SDSS measurements can
be used to explore the seeing dependence on time (on time scales from 1 minute to 10 hours) and 
wavelength (from the UV to the near-IR), as well as its angular correlation on the sky on scales from 
arcminutes to about 2.5 degrees. Thanks to large dynamic range for stellar brightness, the PSF can 
be traced to large radii ($\sim$30 arcsec) and compared to seeing profiles predicted by turbulence
theories\footnote{For most parts of this paper, we consider the SDSS PSF size
same as the seeing, since the PSF is dominated by the atmosphere. However,
the instrument also contributes to the PSF, as discussed in the next Section.}.

The SDSS seeing measurements represent an excellent database that has
not yet been systematically 
explored. Here we utilize about a million SDSS seeing estimates to study the seeing profile 
and its behavior as a function of time and wavelength, and compare our results to theoretical 
expectations. The outline of this paper is as follows. In \S2, we give a brief description of the 
observations and the data used in analysis. We describe the PSF profile analysis, 
including estimation method for the full-width-at-half-maximum (FWHM) seeing
parameter, in \S3. In \S4, we analyse the dependence of FWHM on wavelength, and its angular 
and temporal structure functions. We present and discuss our conclusions in \S5. 

%% file: data_overview.tex

We describe here the SDSS dataset and seeing estimates used in this work. The
selected subset of data, the so-called Stripe 82, represents about one third of
all SDSS imaging data. 

\subsection{Stripe 82 dataset} 

The equatorial Stripe 82 region (22$^h$24$^m$ $<$ R.A. $<$ 04$^h$08$^m$, 
$-$1.27$^\circ$  $<$ Dec $<$ $+$1.27$^\circ$, about 
290 deg$^2$) from the southern Galactic cap ($-64^\circ < b <  -20^\circ$) was repeatedly imaged (of order
one hundred times) by SDSS to study time-domain phenomena (such as supernovae, asteroids, variable stars, quasar 
variability).  An observing stretch of SDSS imaging data is called a ``run''. Often there is only a single
run for a given observing night, though sometimes there are multiple
runs per night. In this paper we use seeing data for 
108 runs, with a total of 947,400 fields, obtained between September,
1998 and September 2008 (there are 6 camera columns, each with 5 filters; for more
details please see \citealt{Gunn2006}). All runs are obtained during the Fall observing season (September to 
December). Astrometric and photometric aspects of this dataset have been discussed in detail by 
\cite{Ivezic2007} and \cite{Sesar2007}.

\subsection{The treatment of seeing in SDSS}
 
Even in the absence of atmospheric inhomogeneities, the SDSS telescope delivers images whose 
FWHMs vary by up to 15\% from one side of a CCD to the other; the worst effects are seen in 
the chips farthest from the optical axis \citep{Gunn2006}. Moreover, since the atmospheric 
seeing varies with time, the delivered image quality is a complex two-dimensional function 
even on the scale of a single field (for an example of the instantaneous image quality across 
the imaging camera, see Figure 7 in \citealt{SDSSEDR}). 
 
The SDSS imaging PSF is modeled 
heuristically in each band using a Karhunen-Lo\'{e}ve (K-L) transform \citep{Lupton2002}. 
Using stars brighter than roughly 20$^{th}$ magnitude, the PSF images from a series of five 
fields are expanded into eigenimages and the first three terms are kept (K-L transform is 
also known as the Principal Component Analysis). The angular variation of the eigencoefficients
is fit with polynomials, using data from the field in question, plus the immediately preceding 
and following half-fields. The success of this K-L expansion is gauged by comparing PSF 
photometry based on the modeled K-L PSFs with large-aperture photometry for the same 
(bright) stars \citep{SDSSEDR}. 
Parameters that characterize seeing for one field of imaging data are stored in the so-called psField 
files\footnote{https://data.sdss.org/datamodel/files/PHOTO\_REDUX/RERUN/RUN/objcs/CAMCOL/psField.html}. 
The status parameter flag for each field indicates the success of the K-L decomposition.

In addition to the K-L decomposition, the SDSS processing pipeline computes parameters of the 
best-fit circular double Gaussian, evaluated at the center of each field. The measured PSF profiles are 
extended to $\sim$30 arcsec using observations of bright stars and at such large radii 
double Gaussian fits underpredict the measured profiles. For this reason, the fits are extended 
to include the so-called ``power-law wings'', which is reminiscent of
the Moffat function,
\begin{equation}
\label{eq:SDSSPSF}
        PSF(r) = {\exp(-{r^2\over 2\,\sigma_1^2}) + b\,\exp(-{r^2\over 2\,\sigma_2^2})
           + p_0\left(1 + { r^2 \over \beta \sigma_P^2}\right)^{-\beta/2} \over 1 + b + p_0}.
\end{equation} 
The measured PSFs are thus modeled using 6 free parameters ($\sigma_1$, $\sigma_2$, $\sigma_P$,
$b$, $p_0$ and $\beta$), and the best-fit parameters are reported in the psField files. 
Given that the measured profiles include only up to 10 data points, the fits are usually excellent
although they do not appear very robust (for examples of bad fits see Fig.~\ref{fig:psffit}).

%% file: PSFprofile.tex

Since the complex 6-parameter PSF fit given by Eq.~(\ref{eq:SDSSPSF}) was adopted by 
the SDSS processing pipeline, significant progress has been made in validating the 
\vk~model of the atmosphere and measuring the associated outer
scale (see, for example, \citealt{Tokovinin2002}, \citealt{Boccas2004}, and \citealt{MartinezMessenger}).
In this section, we describe our much simpler 2-parameter fits to the SDSS PSF
radial profiles using the \vk~atmosphere model.

\begin{figure}[ht]
\centering
\includegraphics[width=0.8\textwidth]{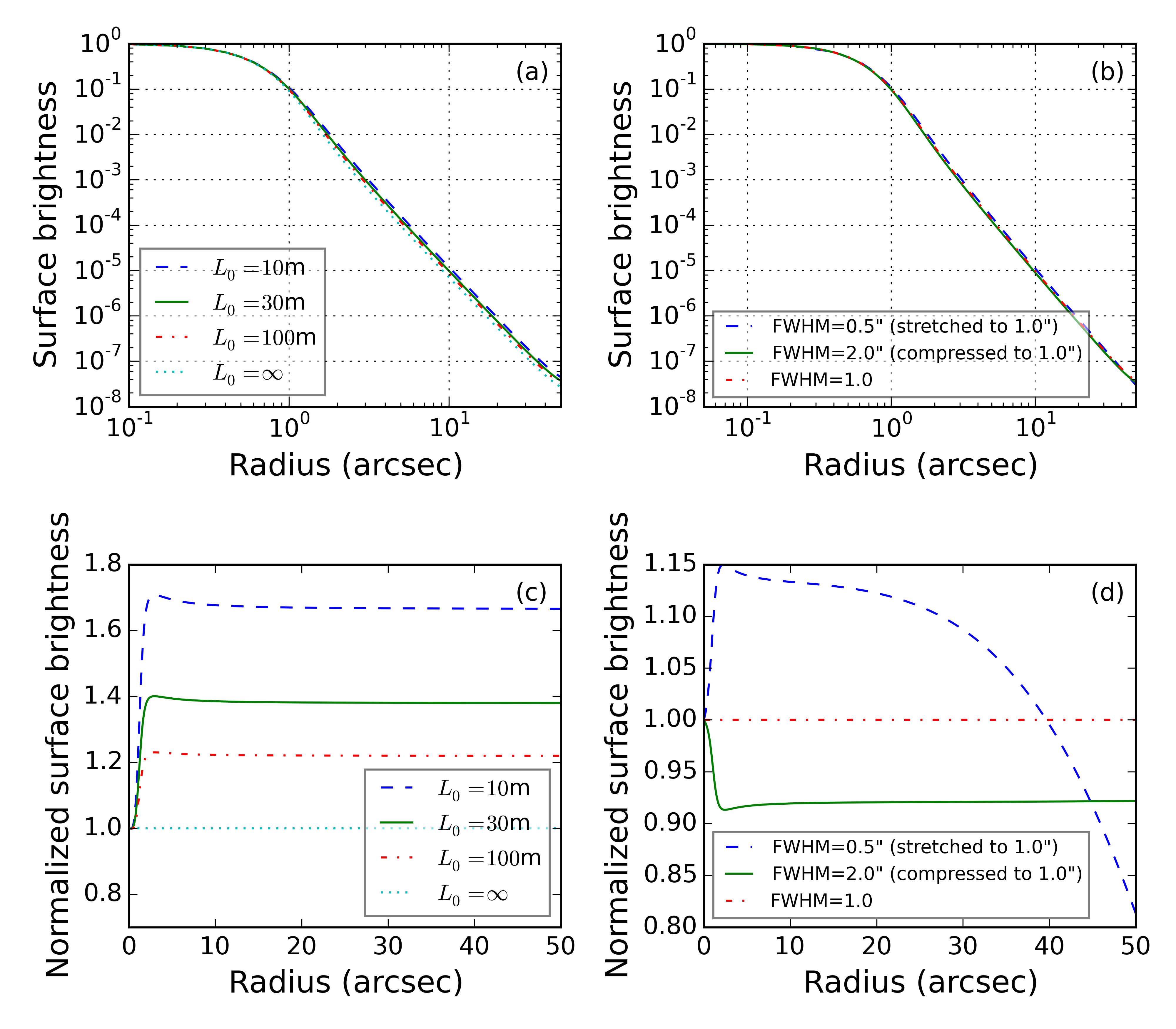}
\vskip -0.2in 
\caption{PSF radial profiles with the \vk~model for a few different
  outer scale ($L_0$) values (left) and $r_0$ values (right). 
All profiles have FWHM of 1.0 arcsec, and
  are normalized to unit peak intensity. The \vk~model becomes
  Kolmogorov when $L_0 = \infty$.
In the right plot, the dashed and dotted profiles are created with
$L_0 = 30$m, and 
FWHM of 0.5 arcsec and 2.0 arcsec, respectively, then stretched and compressed to 1.0 arcsec.
\label{fig:vonK}}
\end{figure}

The seeing profile predicted by the \vk~atmosphere model is a two-parameter
family that can be parametrized by the FWHM and the so-called outer scale
parameter. The Kolmogorov seeing profile is a special case of the 
\vk~seeing with the infinitely large outer scale. The radial profiles of the 
\vk~PSF with a few different values of outer scale are shown in Fig.~\ref{fig:vonK} (a). 
In Fig.~\ref{fig:vonK} (c), the profiles with finite outer scales
($L_0$) are normalized to the Kolmogorov profile.

Our fitting of each PSF radial profile is a 2-step process. First we fit the
measured PSF profile to a \vk~PSF with only one free parameter -
the FWHM of the \vk~profile, while a fiducial outer scale of 30 meters
is assumed fixed. As discernible from Fig.~\ref{fig:vonK} (a) and (c), the impact of the exact
value of $L_0$ on the profile shape is small. A fixed value of $L_0$ induces a small 
systematic uncertainty in the normalization of the contribution of instrumental PSF, 
discussed in \S~\ref{sec:instrPSF}  below. The \vk~PSF profile is generated by creating 
the atmosphere structure function first, as given by Eq. (18) in \cite{Tokovinin2002},
then calculating the PSF through the Optical Transfer Function. 
Pixel integration is taken into account by a factor of 10 oversampling.

Ideally, the Fried parameter $r_0$ would be used as the free
parameter in the fit to the \vk~model. However, 
that would require us to calculate the special functions and do the
Fourier Transform on a large array for each function evaluation.
Instead, we opted to generate a single \vk~PSF template with the FWHM of 
1.0 arcsec. In our one-parameter \vk~fit, we only stretch or compress
the template radially to get the best match with the data, in the
least-square sense.
Fig.~\ref{fig:vonK} (b) shows a comparison of three PSF profiles,
one generated with FWHM of 1.0 arcsec, the other two generated with
FWHM of 0.5 and 2.0 arcsec, respectively, then stretched and
compressed to 1.0 arcsec. The three curves are seen to be almost
indistinguishable.
Fig.~\ref{fig:vonK} (d) shows the same three profiles when normalized
to the one generated with FWHM of 1.0 arcsec.

The $\chi^2$ is defined using the first four data points on the
measured radial profiles, at 0.16, 0.51, 0.87 and 1.44 arcsec,
respectively. These points correspond to highest photon counts, and 
also are least susceptible to errors in the background brightness
estimates. 
The error estimates on these data points come from the original SDSS measurements.

Although the fitted
curves agree with the input data points very well, generally much better than the
original 6-parameter double-Gaussian fit by SDSS, they do not always describe
the PSF tail beyond $\sim 15$ arcsec radius. 
Some examples of such fits are shown in Fig.~\ref{fig:psffit}.
This discrepancy is easily understood
because the PSF tails in the optical bands can be 
different due to the properties of the CCDs.
The SDSS $u$-, $g$-, $r$-, and $z$-bands differ only slightly due to
changing conversion depth. It is known that the $i$-band PSF has ``stronger tails''
because of scattering in the CCD (J.E. Gunn, priv. comm.). The Si is transparent at long $i$-band wavelengths 
so light goes all the way through the chip and is reflected off the solder, and passes 
back up through the Si. This effect is not visible in the $z$-band because in this case
thick front-side illuminated chips are used (in all other bands, thin back-side chips are used).

\subsection{Instrumental PSF \label{sec:instrPSF}} 

\begin{figure}[th]
\centering
\includegraphics[width=1.0\textwidth]{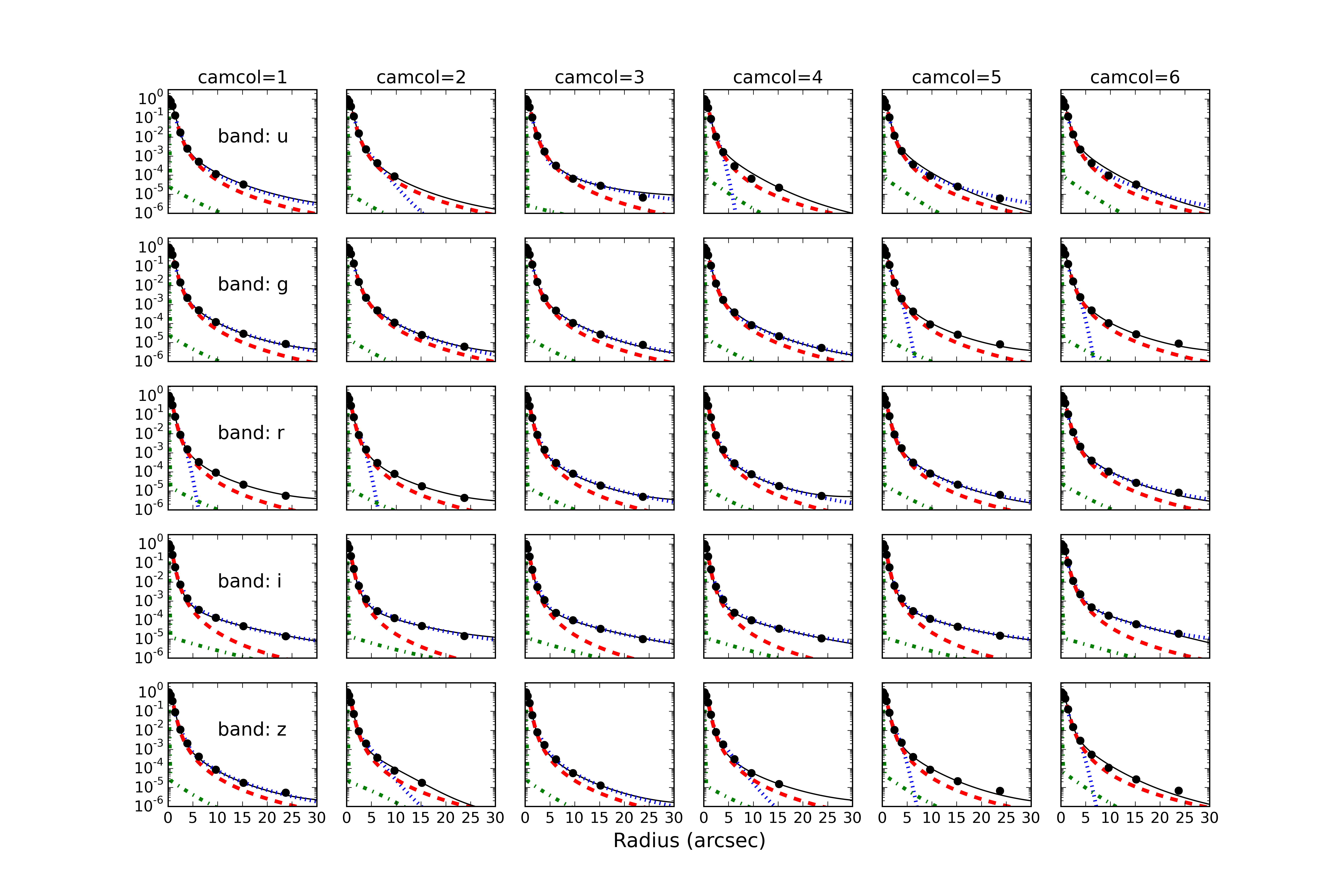}
\vskip -0.3in
\caption{Fits to the PSF radial profiles from run 4874, field 121. Symbols are SDSS data. 
  Red dashed curves are the best one-parameter \vk~fits. Black solid curves are the red
  curve convolved with the instrumental PSF (green dot-dashed lines), where the scaling factor
  (relative normalization) 
  for the tail component is allowed to vary. 
As a reference, the original ``SDSS double Gaussian plus power-law wing'' fits,
described by eq.~\ref{eq:SDSSPSF}, 
 are shown in blue dotted lines -- they sometimes fail catastrophically (see text). 
Note that the y-axis is shown on the logarithmic scale.
\label{fig:psffit}}
\end{figure}

To improve the fit quality at large radii, in the second fitting step we introduce an
empirical ``instrumental'' PSF. Despite the name, this component might also include 
effects not modeled by the \vk~theory, such as aerosol scattering in the atmosphere,
dust on the mirrors, and scattering in the CCDs. The observed PSF can be expressed 
as a convolution of the atmosphere, represented 
by the \vk, and the instrumental PSF,
\begin{equation}
        \textrm{PSF} = \textrm{vonK} (\textrm{FWHM}) \otimes
        \textrm{PSF}_{\textrm{inst}},
\label{eq:conv}
\end{equation} 
where vonK is the \vk~shape, whose only parameter, FWHM, is fixed to
the value from step one.
Fig.~\ref{fig:psffit} shows that the tails of the PSF can be well
described using a second order polynomial in the logarithmic space of
the intensity.
Meanwhile, since $\textrm{PSF}_{\textrm{inst}}$ is a convolution
kernel, we can use a narrow Gaussian to describe its central core.
We define the functional form of the instrumental PSF as
\begin{equation}
        \textrm{PSF}_{\textrm{inst}} = \exp(-\frac{r^2}{2\sigma^2}) + 10^{p(r)},
\label{eq:psfinst}
\end{equation} 
where $p(r)$ is a second order polynomial.
The standard deviation of the Gaussian, $\sigma$, cannot be
too wide because the \vk~term already well describes the core of the
PSF.
We found that $\sigma = 0.1$ arcsec is an acceptable choice.

We define the second order polynomial $p$ as
\begin{equation}
        p(r) = \eta(ar^2+br+1).
\label{eq:psfinstp}
\end{equation} 
Because the shape of the instrumental PSF tail should not vary with
time, but does vary with the filter and camera column,
we determine the values of $a$ and $b$ for each band-camera-column
combination using one representative
field, then fix them at those values for all step-two fits.
For each SDSS PSF radial profile,
these one-time least-square fits use all the data points with radii up
to $\sim$30 arcsec.
Each fit has $a$, $b$, and $\eta$ as free parameters, and involves a 2-dimensional convolution (see Eq.~\ref{eq:conv}),
The fits are very slow but need to be done only once.
We used here run 94, field 11 for these one-time fits, but verified that 
the results are stable for other choices of run and field. 
The best-fit values of $a$ and $b$ are listed in Table~\ref{tab:abc}.

\begin{figure}[th]
\centering
\includegraphics[width=0.5\textwidth]{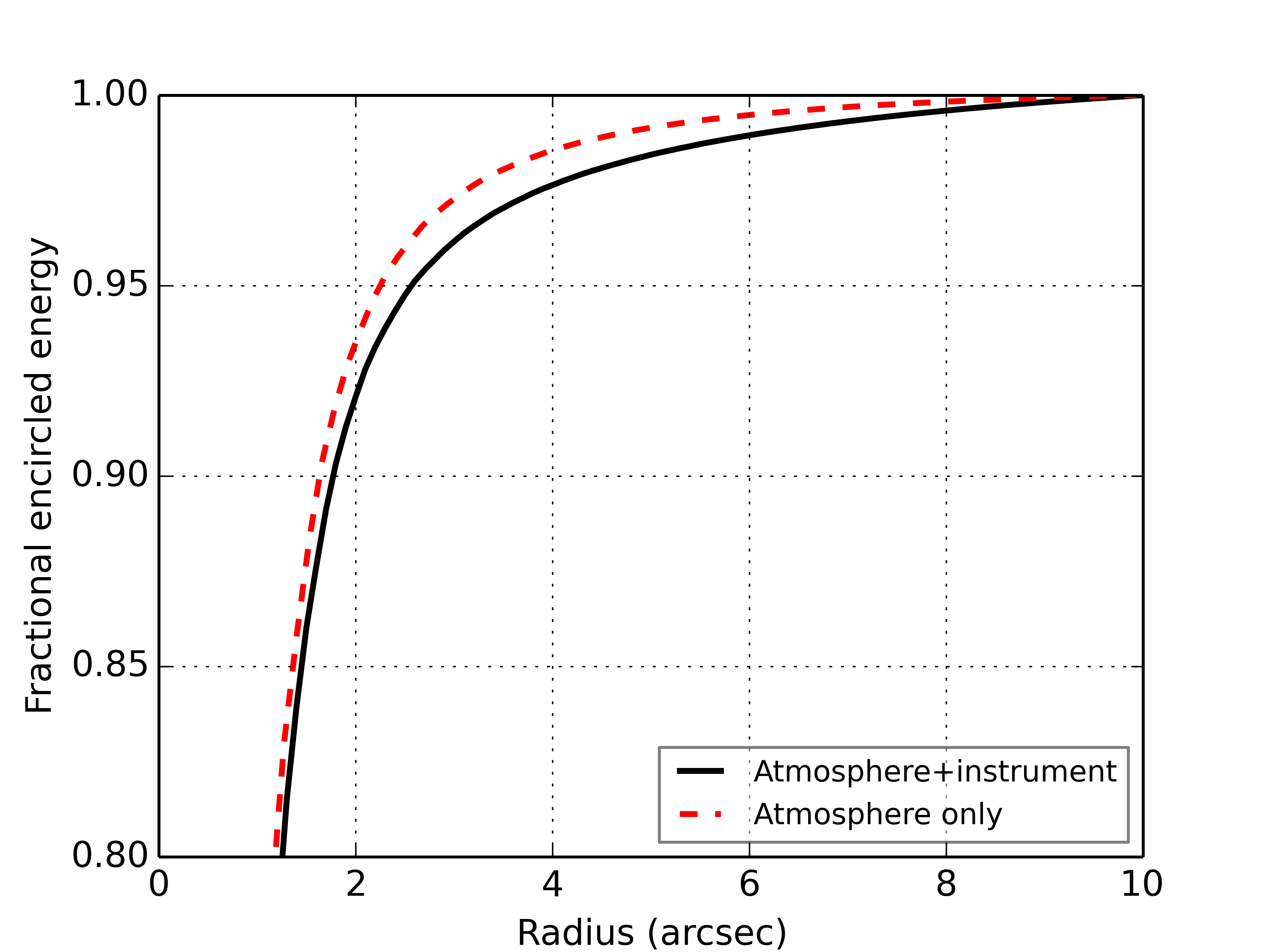}
\caption{Encirlced energy distributions for PSF profiles from run
  4874, field 121, $r$-band, camera column 3. The black solid curve is
  for the delivered PSF with contributions from atmosphere and
  instrument.
The red dashed curve is for atmosphere only.
\label{fig:ee}}
\end{figure}

For step-two PSF fitting, parameters $a$ and $b$ are
fixed for each band-camera-column combination.
$\eta$, the relative normalization of the instrumental PSF
tail in the logarithm-space, is the only free
parameter.
This second fitting step is also a least-square fit with a
2-dimensional convolution, using all the data points
with radii up to $\sim$30 arcsec.
Each two-step PSF fit can be done in a few seconds.
Fig.~\ref{fig:psffit} shows the results of our PSF fits from run 4874. The two-parameter
fits describe the PSF radial profiles quite well, both in the core and
in the tails. The addition of the instrumental PSF 
(dot-dashed lines in Fig.~\ref{fig:psffit})
significantly improves the fit quality, 
especially in the $i$-band. 
Its impact on the encircled energy profile is shown in Fig.~\ref{fig:ee}.

Fig.~\ref{fig:psffit} also shows the original SDSS ``double Gaussian plus power-law wing'' fits,
described by Eq.~\ref{eq:SDSSPSF}. They sometimes fail catastrophically; our analysis revealed
two kinds of failures: one case is characterized by $p_0 =10^{-7}$ and
another by $\beta$=3 or 10.
For the sample of 947,400 PSF fits analyzed here, each failure case occurs with a frequency of about 12\%. 
Inspection of the SDSS code (findPsf.c) reveals that these values signal bad fits which did not 
converge for various (unknown) reasons. 

There are a total of 108 runs in the SDSS Stripe 82 dataset. Among them, run 4874 is the longest, 
with 981 fields. In the rest of this paper, whenever we illustrate results from a single run, 
we always use run 4874 as the fiducial example run.

\begin{table}[th]
\begin{center}
\caption{Values for instrumental PSF shape parameters $a$ and $b$.\label{tab:abc}}
\begin{tabular}{c|c|rrrrrr}
\tableline\tableline
\multicolumn{2}{c|}{} & \multicolumn{6}{c}{Camera Column} \\\cline{3-8}
\multicolumn{2}{c|}{} & 1 & 2 & 3 & 4 & 5 & 6\\\hline
   & a($\times 10^{-4}$) & $-$4.4 & $-$4.4 & $-$1.9 & $-$4.4 & $-$4.4& $-$4.4\\
 $u$& b($\times 10^{-2}$) & 3.3   & 3.3       & 1.3       &      4.7 &4.7   & 4.7\\ \hline
  & a($\times 10^{-4}$) & $-$5.3 & $-$5.3 & $-$4.4 & $-$4.4 & $-$5.3&$-$5.3 \\
 $g$& b($\times 10^{-2}$) & 3.5  & 3.5      & 3.3       &        3.3 &3.5 & 3.5\\\hline
  & a($\times 10^{-4}$) & $-$4.9 & $-$4.9 & $-$4.9 & $-$5.8 &$-$4.4 & $-$4.4\\
 $r$& b($\times 10^{-2}$) & 3.1 & 3.1         & 3.1     &        3.3 &3.3 & 3.3\\\hline
  & a($\times 10^{-4}$) & $-$1.3 & $-$2.2& $-$1.3 & $-$1.3 & $-$1.8& -0.4\\
 $i$& b($\times 10^{-2}$) & 1.7 & 1.8        & 1.7      &        1.7 &1.7 & 1.5\\\hline
  & a($\times 10^{-4}$) & $-$6.2 & 4.4     & $-$6.2 & $-$4.4  &$-$6.2 & $-$4.4\\
 $z$& b($\times 10^{-2}$) & 4.0 & 2.0      & 4.0      &        3.1 &4.4 & 4.7\\
\tableline
\end{tabular}
\end{center}
\end{table}

%% file: FWHManalysis.tex

Given that the observed seeing is by and large described by a single parameter, FWHM, 
we study here three aspects of its variation in detail: dependence on wavelength,
the spatial (angular) structure function, and temporal behavior. We note that details 
about the seeing profile tails, including the contribution of the instrumental profile and
the $i$ band behavior, do not matter here because we focus only on the FWHM behavior. 

\subsection{The FWHM dependence on wavelength} 

The Kolmogorov turbulence theory gives a standard formula for the FWHM of a long-exposure
seeing-limited PSF in a large telescope,
\begin{equation}
\textrm{FWHM}^{\rm Kolm}(\lambda, X) = \frac{0.976\lambda}{r_0(\lambda,X)},
\label{eq:fwhmkolm}
\end{equation}
\begin{equation}
r_0(\lambda,X) = r_0(\lambda_0, 1) \left(\frac{\lambda}{\lambda_0}\right)^{1.2}
\frac{1}{X^{0.6}},
\label{eq:r0}
\end{equation}
where $\lambda$ is the wavelength in meter, $X$ is the airmass,
$r_0$ is the Fried parameter in meter, and $\textrm{FWHM}^{\rm Kolm}$
is in radian.
We use $\lambda_0$ as the reference wavelength.
$r_0(\lambda_0, 1)$ is the $r_0$ for $\lambda=\lambda_0$ and $X$=1.
Substituting Eq.~(\ref{eq:r0}) into (\ref{eq:fwhmkolm}), it is easy to show that 
\begin{equation}
\textrm{FWHM}^{\rm Kolm} \propto \lambda^{-0.2}.
\end{equation}

With the \vk~atmosphere model, the FWHM as in
Eq.~(\ref{eq:fwhmkolm}) needs an additional correction factor
which is a function of the outer scale $L_0$~\citep{Tokovinin2002},
\begin{equation}
\label{eq:FWHMvK}
\textrm{FWHM}^{\rm vonK}(\lambda, X) = \frac{0.976\lambda}{r_0(\lambda,X)}
\sqrt{1-2.183\left( \frac{r_0(\lambda,X) }{L_0} \right)^{0.356}}.
\end{equation}
If a power-law approximation is attempted,  
\begin{equation}
\textrm{FWHM}^{\rm vonK} \propto \lambda^{\alpha},
\label{eq:fwhmvonk} 
\end{equation}
$\alpha$ becomes a function of $L_0$ and $r_0$ at a specified
wavelength and airmass, or equivalently, a function of $L_0$ and FWHM$^{\rm vonK}$.
For the subsequent analysis, we adopt the $r$ band as the fiducial band (with
the effective wavelength of 616.6 nm).

\begin{figure}[th]
\centering
\includegraphics[width=0.9\textwidth]{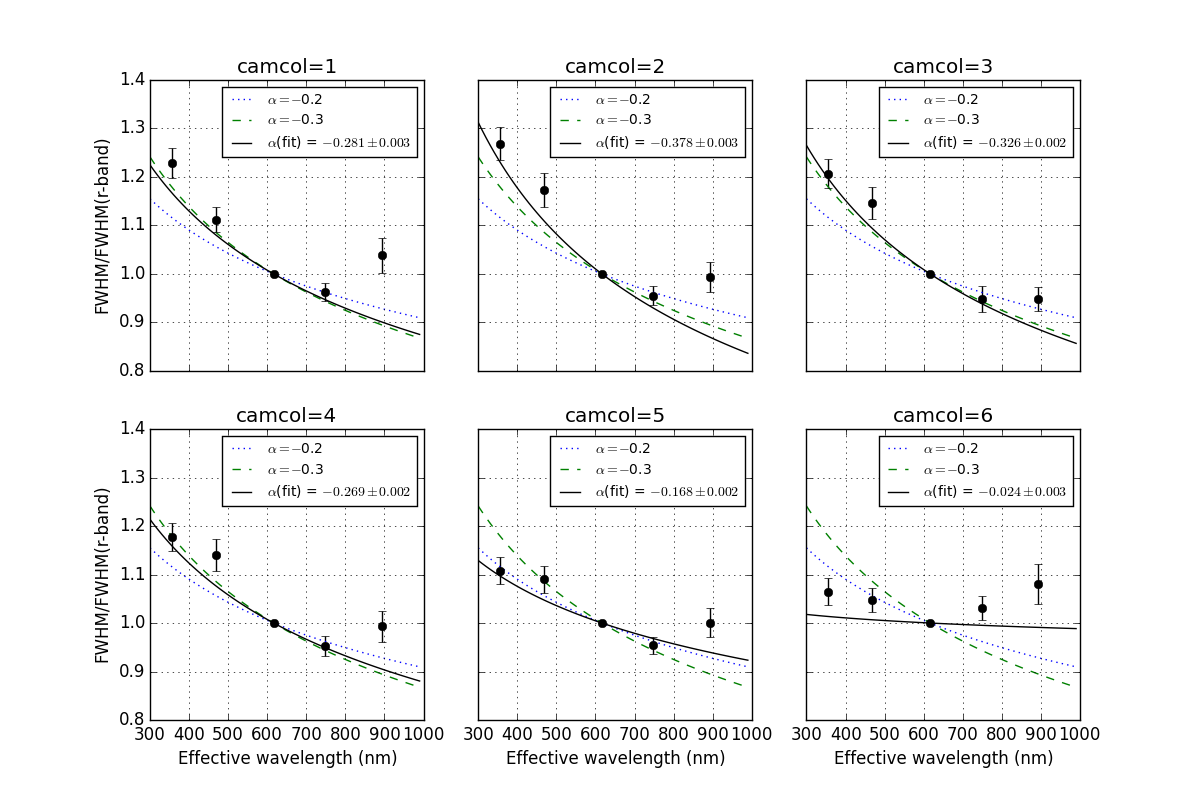}
\caption{The behavior of FWHM as a function of wavelength for the fiducial run 4874.
Symbols are SDSS data and solid line is the best power-law fit, with the best-fit slope
($\alpha$) shown in inset. For comparison purposes, the $\alpha=-0.2$ (dotted) and $\alpha=-0.3$ 
(dashed) lines are also shown. For the ensemble behavior of best-fit $\alpha$, see Fig.~\ref{fig:alpha_fwhm}. 
\label{fig:fwhm_lambda}}
\end{figure}

\begin{figure}[th]
\centering
\includegraphics[width=0.5\textwidth]{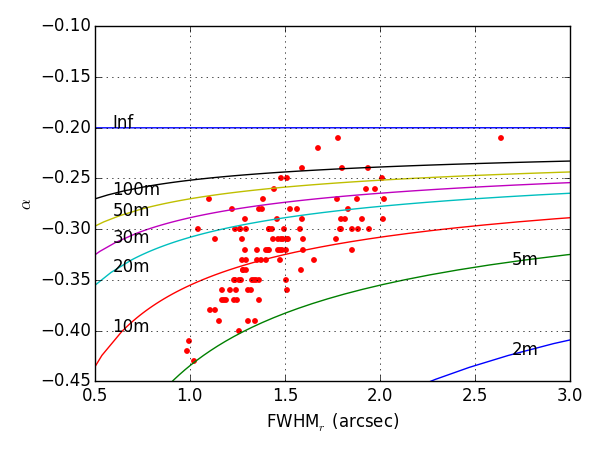}
\caption{The variation of the best-fit power-law index for the wavelength dependence of FWHM, $\alpha$, 
vs. the FWHM in the $r$-band for all the 108 Stripe 82 runs. The symbols are SDSS
measurements of $\alpha$ based on the $ugri$ data and averaged over camera columns 2 to 5. 
The curves are predictions of the 
\vk~model, with $L_0$ ranging from 2 meters to infinity, as labeled. The data are clearly
inconsistent with Kolmogorov predictions ($L_0=\infty$) and reasonably well described by
\vk~model and $L_0$ in the range from 5m to $\sim$100m.  \label{fig:alpha_fwhm}}
\end{figure}

For each run from SDSS Stripe 82 data, and each camera column, we make
a least-square fit to all the simultaneous FWHM measurements across the optical bands, to
estimate the power-law index $\alpha$ (see Eq.~\ref{eq:fwhmvonk}). 
The errors on the FWHM measurements in each optical band comes from
averaging over all the fields.
All FWHM values are multiplied by $1/X^{0.6}$ to 
correct for the airmass effects\footnote{The airmass dependence for the \vk~model 
is not strictly a power law, but can be approximated by a power law with good precision.
By numerically fitting a power law to eq.~\ref{eq:FWHMvK}, we obtained a power-law index 
of 0.63. We ignore the difference between 0.63 and 0.6 as it results in seeing variations 
below 1\% for the probed range of airmass.}.

All FWHM are normalized using 
corresponding FWHM in the $r$-band taken at the same moment in time. 
We take into account that the same field number does not correspond to the same
time in all filters. The scanning order in the SDSS camera is $r$-$i$-$u$-$z$-$g$, with the delay between the two 
successive filters corresponding to 2 fields. That is, if we take the field number $F$ for the $r$-band, then
we need to take FWHM for the $i$-band from field $F-2$, for the $u$-band
from $F-4$, and so on. 

Fig.~\ref{fig:fwhm_lambda} shows such fits for run 4874. Significant deviation 
from $\alpha = -0.2$, predicted by the Kolmogorov model, can be seen in most bands.
We find that fits in columns 1-5 are always similar, while in column 6 the slope is 
systematically lower. Similarly, the data in the $ugri$ bands are well fit by the power law, 
while the $z$ band the data are systematically larger than the power-law fit. 
For this reason, we refit the data using only $ugri$ bands and average results without
using the edge columns (1 and 6, though including column 1 does not substantially change 
the results). Fig.~\ref{fig:alpha_fwhm} shows a scatter plot of the resulting
best-fit $\alpha$ vs. the FWHM in the $r$-band, for all the analyzed runs.
 
As discussed above, according to the \vk~atmosphere model, the
power index $\alpha$ should be a function of the outer scale $L_0$ and 
FWHM. A correlation between $\alpha$ and the FWHM is discernible in
Fig.~\ref{fig:alpha_fwhm}. Similar correlations have been seen in Subaru images 
and reported by~\cite{subaruSeeing2016}.
The data points are overlaid with curves predicted by the 
\vk~model, with $L_0$ varying from 2 m to infinity.
The data clearly deviate from the Kolmogorov model prediction, which is
the horizontal line at $\alpha = -0.20$, with an infinite $L_0$.
For example, for LSST's fiducial FWHM of 0.6 arcsec and the commonly assumed 
$L_0 = 30$ m, the \vk~model predicts an $\alpha$ value close to $-0.31$.

\subsection{Angular structure function} 

\begin{figure}[th]
\centering
\includegraphics[width=0.5\textwidth]{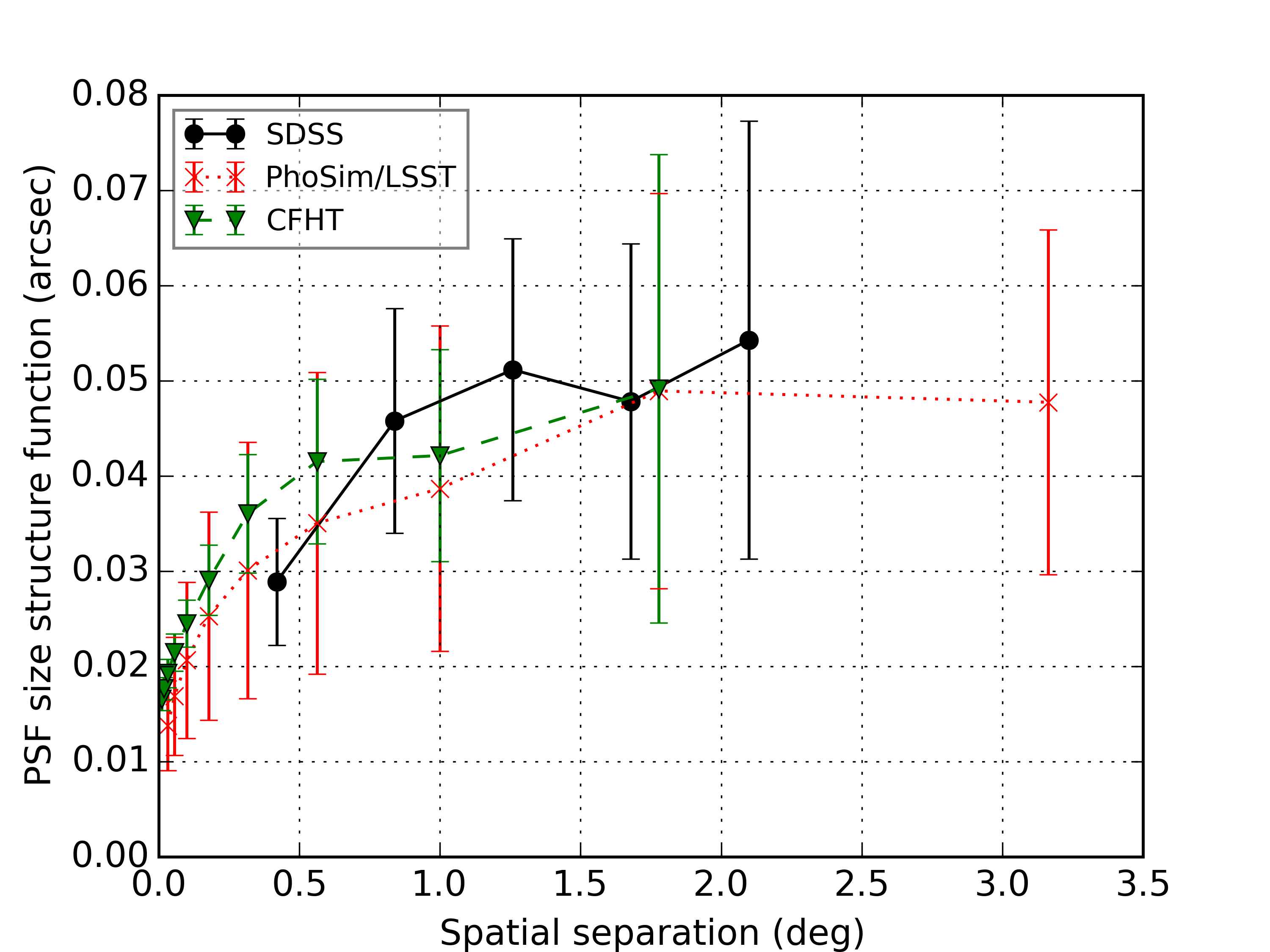}
\caption{The angular structure function for the PSF size determined using 
  CFHT data from \cite{heymans2012}, SDSS data analyzed here, and LSST image simulations. 
  SDSS measurements are averaged over 86 runs with the number of fields larger than 100. 
\label{fig:spatial}}
\end{figure}

To examine the angular (spatial) correlation of the FWHM, we compute the angular
structure function using PSF measurements from all 6 camera columns.
Our structure function is defined as the root-mean-square scatter of the PSF size 
differences of pairs of stars in the same distance bin along the direction perpendicular
to the scanning direction\footnote{The adopted form 
of the structure function, $SF$, is closely related to the autocorrelation function, $ACF$, as 
$SF \propto (1-ACF)^{1/2}$.} .
The SDSS curves are combined for 86 Stripe 82 runs with the number of fields larger than 
100 (out of 108 runs) 
We also compared the structure functions for each band
separately, with and without camera columns 1 and 6, 
and found no statistically significant differences.
Results for the $r$-band are shown in Fig.~\ref{fig:spatial}.

The structure function starts saturating at separations of
$\sim 0.5 - 1.0$ degree, with an asymptotic value of about $\sim 0.05$ arcsec.
In other words, the seeing rms variation at large angular scales is about 5\%,
but we emphasize that our data do not probe scales beyond 2.5 degree. 

For comparison, Fig. ~\ref{fig:spatial} also shows results from the CFHT PSF 
measurements~\citep{heymans2012}, and simulated PSF angular
structure functions obtained using image simulation code PhoSim~\citep{phosim}. 
The PhoSim PSF profiles are obtained by simulating a grid of stars
spaced by 6 arcminutes with non-perturbed LSST telescope and ideal sensors.
The results are averaged over 9 different atmosphere realizations with
different wind and screen parameters and airmass, and over 3 different
wavelengths (350 nm, 660 nm, and 970 nm).
The CFHT PSF size measurements were made in the $i$-band, and provided
by the authors of~\cite{heymans2012}.
The three curves in Fig.~\ref{fig:spatial} appear to be quantitatively
consistent with each other, even though they correspond to telescopes at
different sites and with different optics. 

We note that the PhoSim code could be used to further quantitatively study the variation
of seeing with outer scale and the impact of telescope diameter; however,
such detailed modeling studies are beyond the scope of this report.

\subsection{Temporal behavior}

\subsubsection{Power spectrum analysis} 

\begin{figure}[th]
\centering
\includegraphics[width=0.99\textwidth]{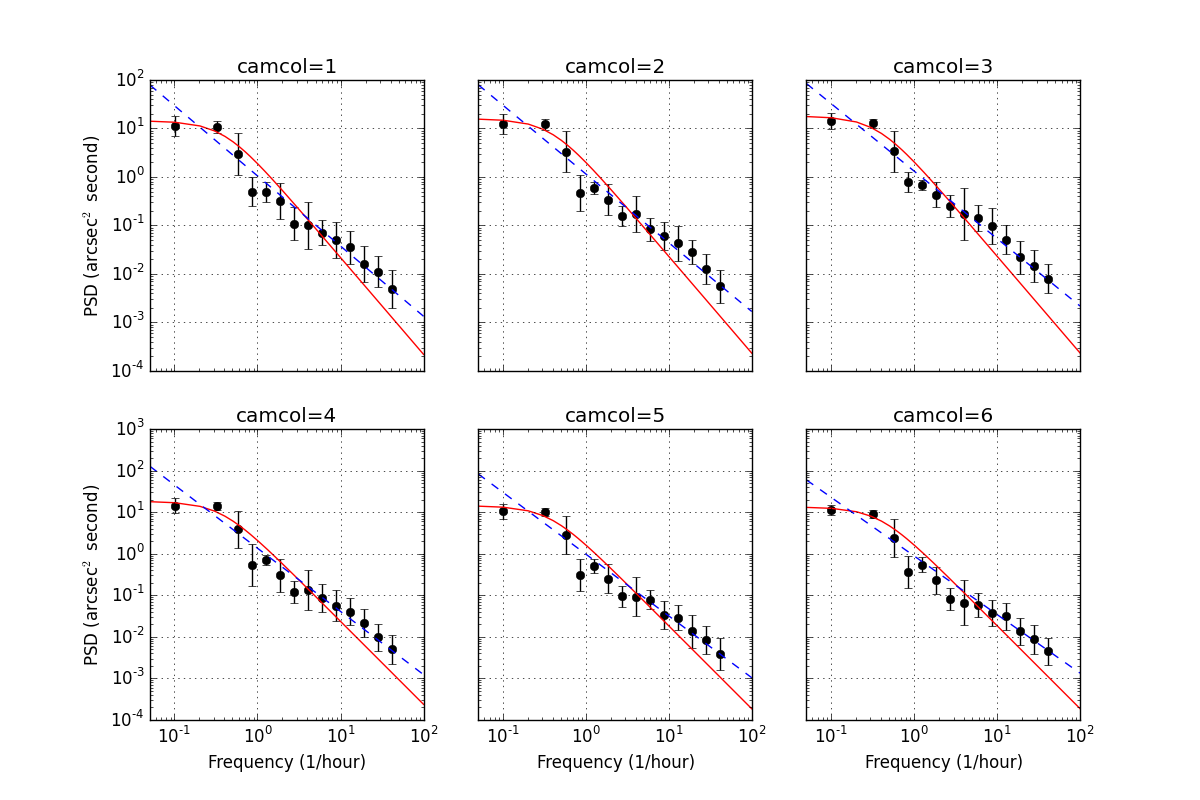}
\vskip -0.2in
\caption{PSF size temporal power spectral density for run 4874, r-band. 
The solid lines are fits using the damped random walk model. 
The dashed lines show best fits based on a single power law. The former
predicts a steeper high-frequency behavior, while the latter cannnot 
explain the turnover at low frequencies. 
\label{fig:psd}}
\end{figure}

To study the temporal behavior of the seeing, we first analyze its power spectrum.
Fig.~\ref{fig:psd} shows the temporal power spectral density (PSD) of the
PSF FWHM for 6 camera columns, in run 4874, $r$-band.
The time difference between subsequent fields is 36 seconds. 
Even though anormalies on the wavelength dependence of the FWHM are
seen in column 6, it is clear from Fig. 6 that the temporal behavior
of the FWHM does not vary with band. The temporal analysis presented
in this section includes all the camera columns and all the optical
bands. We have repeated the analysis without using FWHM measurements
from $z$-band and camera columns 1 and 6, and found no statistically significant
differences in results.

\begin{figure}[th]
\centering
\includegraphics[width=0.7\textwidth]{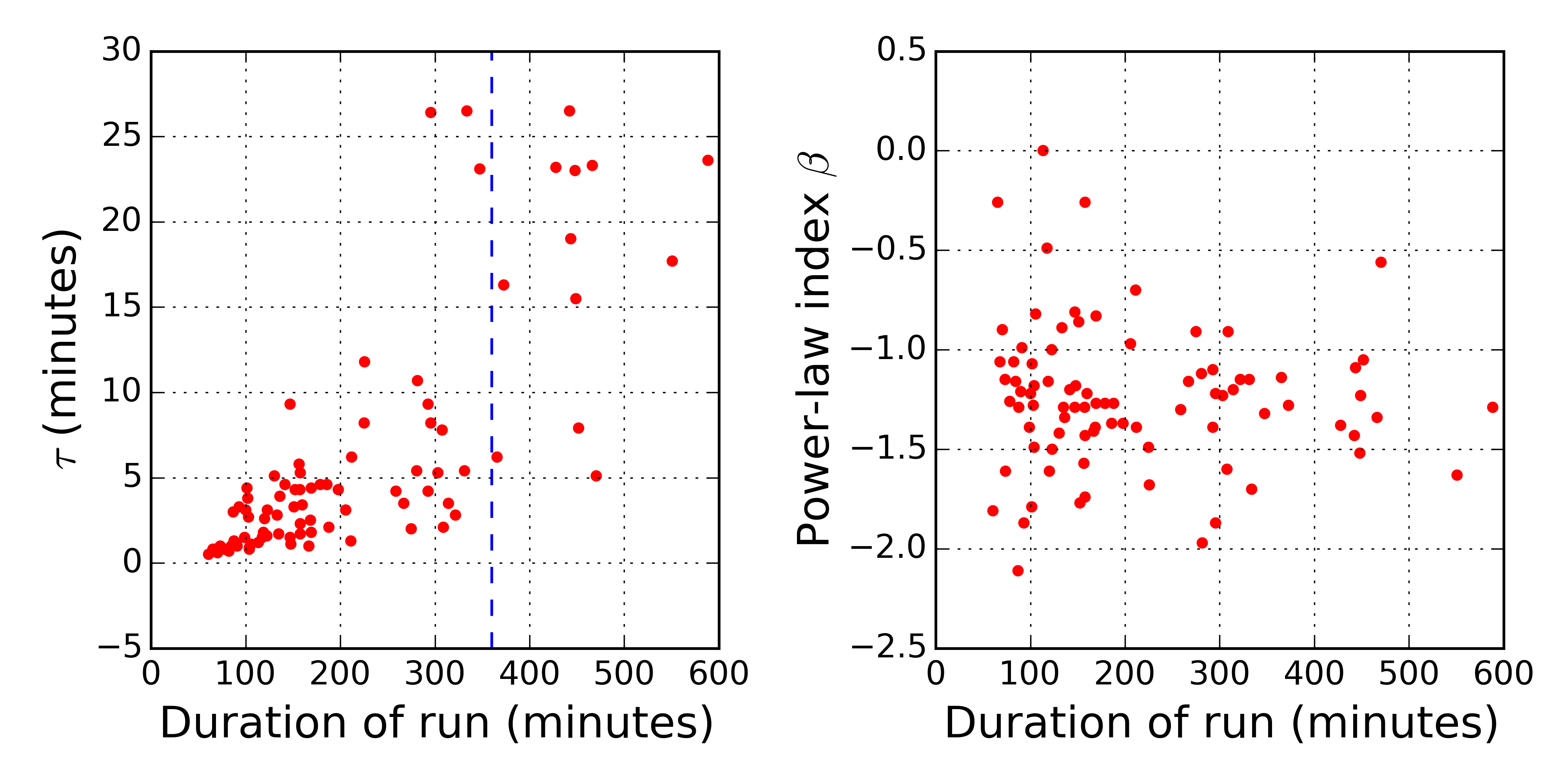}
\vskip -0.2in
\caption{Left: The symbols show the best-fit characteristic timescale $\tau$ in 
damped random walk for all 108 runs in Stripe 82 vs. the duration of 
each run. It is plausible that short runs cannot reliably constrain
$\tau$.
The vertical dashed line indicates run duration of 6 hours.
Right: The power-law index $\beta$ for a single power law fit for all 108 
runs vs. the duration of each run. Note that for the majority of runs, $\beta$
is larger than the value appropriate for damped random walk ($\beta = -2$). 
\label{fig:hist}}
\end{figure}

We fit the PSD using two competing models.
The first is a damped random walk (DRW) model~\citep[for introduction see Chapter 10 in][]{zeljkoBook},
\begin{equation}
\textrm{PSD}(f) = \frac{\tau^2 SF^2_{\infty}}{1+(2\pi f \tau)^2},
\end{equation}
where $f$ is the temporal frequency, $SF_{\infty}$ is the asymptotic
value of the structure function, and $\tau$ is the
characteristic timescale.
The solid curves in Fig.~\ref{fig:psd} show fits using this model,
with $f$, $\tau$, and $SF_{\infty}$ as free parameters.
Note that due to the lack of data toward the low-frequency end, the
first and second bins are four and two times wider than the
rest of the bins, respectively.
Combining fit results for all camera columns and optical bands for run 4874
gives $\tau = 23.6 \pm 1.3$ minutes.
Making the same fits for all the 108 runs in Stripe 82, 
we obtain the $\tau$ distribution vs. the duration of each
run, as shown in Fig.~\ref{fig:hist} (left).
The shorter runs tend to give smaller timescale. It is plausible that short runs 
cannot reliably constrain $\tau$ due to the lack of data toward the low-frequency 
end of the spectra. There are 12 runs longer than 6 hours and their characteristic timescales
are within the range of about $\sim5-30$ minutes.
This result is generally consistent with ~\cite{Racine1996}, where a timescale of 
$\tau = 17 \pm 1$ minutes was found.

The data consistently show a shallower high-frequency behavior than predicted
by damped random walk ($\propto 1/f^2$). In order to quantitatively describe 
the high-frequency tail of the PSD, we fit a simple power law,
\begin{equation}
\textrm{PSD}(f) = B f^\beta,
\end{equation}
where $B$ is the normalization factor, and $\beta$ is the power-law index.
Best-fits are illustrated for run 4874 are in Fig.~\ref{fig:psd} (dashed lines).
Combining fit results for all camera columns and filters gives $\beta = -1.29\pm 0.09$ 
for run 4874. Making the same fits for all the 108 runs in Stripe 82, we obtained the 
$\beta$ distribution vs. the duration of each run shown in Fig.~\ref{fig:hist} (right).
The shorter runs give $\beta$ values with a larger variance, but nevertheless it is 
clear that for most runs the high-frequency behavior can be described with a 
power-law index in the range $-1.5$ to $-1.0$. 
We note that \cite{2016SPIE.9906E..42S} measured steeper slopes,
though at much higher frequencies.
On the other hand, a single 
power law cannot explain the turnover at low frequencies. 

Therefore, neither model provides a satifactory fit over the entire frequency range: 
the power law fit systematically over predicts the low-frequency part of the PSD,
while the $1/f^2$ high-frequency behavior of damped random walk model is too 
steep. It is likely that a hybrid model would work, for example, a simple
generalization of the random walk model
(see Eq. (19) in \cite{Dunkley2005}), 
\begin{equation}
\textrm{PSD}(f) = \frac{\tau^\alpha SF^2_{\infty}}{1+(2\pi f
  \tau)^\alpha}.
\label{eq:hybrid}
\end{equation}
Performing fits to our data using this model did not yield useful
results on the characteristic timescale, due to our lack of data
points at low frequency, and therefore the incapability in
constraining one additional parameter ($\alpha$ in Eq.~(\ref{eq:hybrid})).
We leave more detailed analysis,
perhaps informed by the PhoSim modeling, for future work.

\subsubsection{Structure function analysis}

\begin{figure}[th]
\centering
\includegraphics[width=0.7\textwidth]{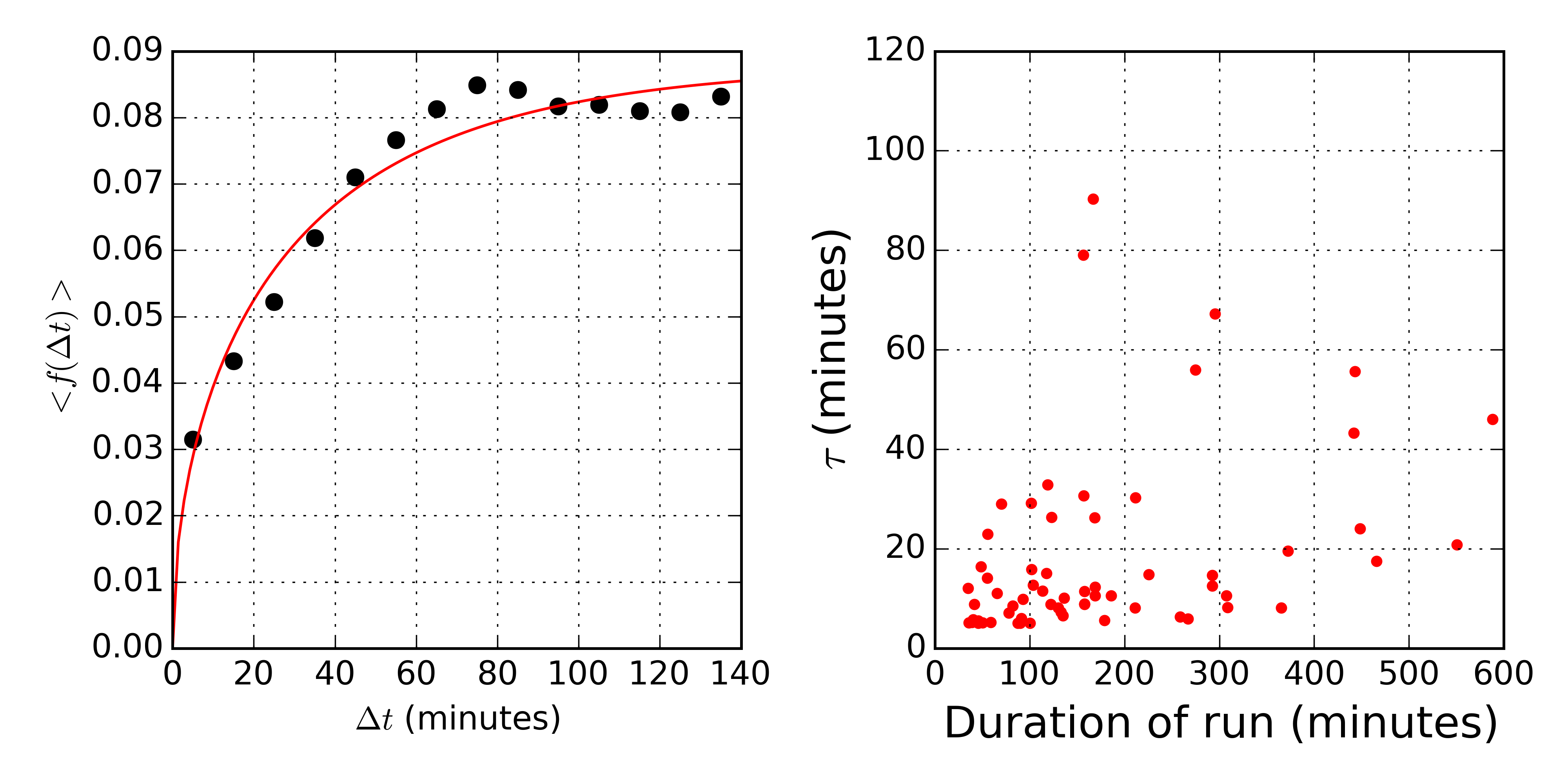}
\caption{Left: The average normalized seeing difference, $<f(\Delta t)>$, as
  a function of the time separation, $\Delta t$, for run 4874, camera
   column 1, in the $r$-band. The fit to Eq.~(\ref{eq:fdt}) gives $f(\Delta t) ^\infty =  
   0.088\pm0.005$, $\tau = 45.1\pm10.3$ minutes and $\gamma$ =
   1.016$\pm$0.102. 
A damped random walk model has $\gamma=1$.
Right: The timescale $\tau$ vs. the duration of each run.
 Note that for most runs $\tau$ is in the range 5-30 minutes
    (runs shorter than 20 minutes or where the fitted $\tau$ is longer than 2/3 of the duration
    of the run are left out).  
\label{fig:fdt}}
\end{figure}

An alternative approach to power spectrum analysis is offered by auto-correlation 
and structure function analysis. Following~\cite{Racine1996}, we define a 
structure-function-like quantity
\begin{equation}
       f(\Delta t) = {| \theta(t+\Delta t) - \theta(t)| \over  \theta(t+\Delta t) + \theta(t) },
\end{equation} 
where $\theta$ is seeing. We then fit the mean value of $f(\Delta t)$ to 
\begin{equation}
    < f(\Delta t) > =  f(\Delta t) ^\infty \, \left( 1 - \exp[-(\Delta
      t/\tau)^\gamma] \right)^{1/2},
\label{eq:fdt}
\end{equation} 
with $f(\Delta t) ^\infty$, $\tau$ and $\gamma$ as free parameters.
Fig.~\ref{fig:fdt} (left) shows one example of such fits. This functional form is 
somewhat inspired by damped random walk model, where $\gamma=1$ and the 
term in brackets is raised to the power of a half. 
For this particular fit in Fig.~\ref{fig:fdt} (left), $\gamma$ is very close to 1.

The best-fit $\gamma$ is found to be mostly in the range 0.5 -1.5. 
The distribution of $\tau$ vs. 
the duration of each run is shown in Fig.~\ref{fig:fdt} (right) (somewhat 
arbitrarily,
runs shorter than 20 minutes and those where the fitted $\tau$ is longer than 2/3 of the duration
    of the run are deemed unreliable and not shown).  
It is evident that 
for most runs the timescale $\tau$ is in the range 5-30 minutes.
Therefore, this analysis seems more robust at constraining the characteristic time
scale than fitting a damped random walk model to empirical PSD.

%% file: conclusions.tex
 
The atmospheric seeing due to atmospheric turbulence plays a major role in 
ground-based astronomy; seeing varies with the wavelength of observation
and with time, on time scales ranging from minutes to years. Better empirical
and theoretical understanding of the seeing behavior can inform optimization 
of large survey programs, such as LSST. We have utilized here for the first 
time an unprecedentedly large database of about a million SDSS seeing estimates
and studied the seeing profile and its behavior as a function of time and wavelength.

We find that the observed PSF radial profile can be parametrized by only two parameters, 
the FWHM of a theoretically motivated profile and a normalization of the contribution 
of an empirically determined instrumental PSF. The profile shape is very similar to 
the ``double gaussian plus power-law wing'' decomposition used by SDSS image
processing pipeline, but here it is modeled with two free model parameters, rather 
than six as in SDSS pipeline (of course, the SDSS image processing pipeline had
to be designed with adequate flexibility to be able to efficiently and robustly 
handle various unanticipated behavior). We find that the PSF radial profile is well 
described by theoretical predictions based on both 
Kolmogorov and \vk's turbulence theory (see Fig.~\ref{fig:psffit}). Given the extra 
degree of freedom due to the instrumental PSF, the shape of the
measured radial profile alone 
is insufficient to reliably rule out either of the two theoretical profiles.  

We report empirical evidence that the wavelength dependence of the atmospheric 
seeing and its correlation with the seeing itself agrees better with the \vk~model 
than the Kolmogorov turbulence theory (see Fig.~\ref{fig:alpha_fwhm}). For example, 
the best-fit power-law index to describe the seeing wavelength dependence in conditions 
representative for LSST survey is much closer to $-0.3$ than to the usually assumed value
of $-0.2$ predicted by the Kolmogorov theory. 
We note that most of the long-term seeing statistics are measured at visible 
wavelengths. The knowledge of the wavelength-dependence of the seeing is useful 
for extrapolating the seeing statistics to other wavelengths, for example, to
the near-infrared where a lot of the adaptive optics programs operate. 
PSF-sensitive galaxy measurements require that the PSFs measured from
stars be interpolated both spatially and in color into galactic PSFs.

We have also measured the characteristic angular and temporal scales on which 
the seeing decorrelates. The angular structure function saturates at scales beyond 
0.5$-$1.0 degree. The seeing rms variation at large angular scales is about 5\%,
but we emphasize that our data do not probe scales beyond 2.5 degree. Comparisons 
with simulations of the LSST and PSF measurements at the CFHT site show good general 
agreement.

The power spectrum of the temporal behavior is found to be broadly consistent with 
a damped random walk model with characteristic timescale in the range $\sim5-30$ 
minutes, though data show a shallower high-frequency behavior. The high-frequency 
behavior can be quantitatively described by a single power law with index in the 
range $-1.5$ to $-1.0$. A hybrid model is likely needed to fully capture both the 
low-frequency and high-frequency behavior of the temporal variations of atmospheric
seeing. 

We conclude by noting that, while our numerical results may only apply to the SDSS 
site, they can be used as useful reference points when considering spatial and 
temporal variations of seeing at other observatories.

%% file: sdsspsf.bbl
\begin{thebibliography}{}
\expandafter\ifx\csname natexlab\endcsname\relax\def\natexlab#1{#1}\fi
\providecommand{\url}[1]{\href{#1}{#1}}
\providecommand{\dodoi}[1]{doi:~\href{http://doi.org/#1}{\nolinkurl{#1}}}
\providecommand{\doeprint}[1]{\href{http://ascl.net/#1}{\nolinkurl{http://ascl.net/#1}}}
\providecommand{\doarXiv}[1]{\href{https://arxiv.org/abs/#1}{\nolinkurl{https://arxiv.org/abs/#1}}}

\bibitem[{{Boccas}(2004)}]{Boccas2004}
{Boccas}, M. 2004, Gemini Technical Note

\bibitem[{{Borgnino}(1990)}]{vk1}
{Borgnino}, J. 1990, \ao, 29, 1863, \dodoi{10.1364/AO.29.001863}

\bibitem[{{Dunkley} {et~al.}(2005){Dunkley}, {Bucher}, {Ferreira}, {Moodley},
  \& {Skordis}}]{Dunkley2005}
{Dunkley}, J., {Bucher}, M., {Ferreira}, P.~G., {Moodley}, K., \& {Skordis}, C.
  2005, \mnras, 356, 925, \dodoi{10.1111/j.1365-2966.2004.08464.x}

\bibitem[{{Gunn} {et~al.}(2006){Gunn}, {Siegmund}, {Mannery}, {Owen}, {Hull},
  {Leger}, {Carey}, {Knapp}, {York}, {Boroski}, {Kent}, {Lupton}, {Rockosi},
  {Evans}, {Waddell}, {Anderson}, {Annis}, {Barentine}, {Bartoszek}, {Bastian},
  {Bracker}, {Brewington}, {Briegel}, {Brinkmann}, {Brown}, {Carr},
  {Czarapata}, {Drennan}, {Dombeck}, {Federwitz}, {Gillespie}, {Gonzales},
  {Hansen}, {Harvanek}, {Hayes}, {Jordan}, {Kinney}, {Klaene}, {Kleinman},
  {Kron}, {Kresinski}, {Lee}, {Limmongkol}, {Lindenmeyer}, {Long}, {Loomis},
  {McGehee}, {Mantsch}, {Neilsen}, {Neswold}, {Newman}, {Nitta}, {Peoples},
  {Pier}, {Prieto}, {Prosapio}, {Rivetta}, {Schneider}, {Snedden}, \&
  {Wang}}]{Gunn2006}
{Gunn}, J.~E., {Siegmund}, W.~A., {Mannery}, E.~J., {et~al.} 2006, \aj, 131,
  2332, \dodoi{10.1086/500975}

\bibitem[{{Heymans} {et~al.}(2012){Heymans}, {Rowe}, {Hoekstra}, {Miller},
  {Erben}, {Kitching}, \& {van Waerbeke}}]{heymans2012}
{Heymans}, C., {Rowe}, B., {Hoekstra}, H., {et~al.} 2012, \mnras, 421, 381,
  \dodoi{10.1111/j.1365-2966.2011.20312.x}

\bibitem[{{Ivezi{\'c}} {et~al.}(2014){Ivezi{\'c}}, {Connolly}, {VanderPlas}, \&
  {Gray}}]{zeljkoBook}
{Ivezi{\'c}}, {\v Z}., {Connolly}, A.~J., {VanderPlas}, J.~T., \& {Gray}, A.
  2014, {Statistics, Data Mining, and Machine Learning in Astronomy}

\bibitem[{{Ivezi{\'c}} {et~al.}(2007){Ivezi{\'c}}, {Smith}, {Miknaitis}, {Lin},
  {Tucker}, {Lupton}, {Gunn}, {Knapp}, {Strauss}, {Sesar}, {Doi}, {Tanaka},
  {Fukugita}, {Holtzman}, {Kent}, {Yanny}, {Schlegel}, {Finkbeiner},
  {Padmanabhan}, {Rockosi}, {Juri{\'c}}, {Bond}, {Lee}, {Stoughton}, {Jester},
  {Harris}, {Harding}, {Morrison}, {Brinkmann}, {Schneider}, \&
  {York}}]{Ivezic2007}
{Ivezi{\'c}}, {\v Z}., {Smith}, J.~A., {Miknaitis}, G., {et~al.} 2007, \aj,
  134, 973, \dodoi{10.1086/519976}

\bibitem[{{Ivezi\'c} {et~al.}(2008){Ivezi\'c}, Tyson, Acosta, Allsman,
  Anderson, Andrew, Angel, Axelrod, Barr, Becker, {et~al.}}]{LSSToverview}
{Ivezi\'c}, {\v{Z}}., Tyson, J.~A., Acosta, E., {et~al.} 2008,
  arXiv:0805.2366v4.
\newblock \doarXiv{0805.2366v4}

\bibitem[{{Lupton} {et~al.}(2001){Lupton}, {Gunn}, {{Ivezi{\'c}}, \v{Z}.},
  {Knapp}, \& {Kent}}]{Lupton2001}
{Lupton}, R., {Gunn}, J.~E., {{Ivezi{\'c}}, \v{Z}.}, {Knapp}, G.~R., \& {Kent},
  S. 2001, in Astronomical Society of the Pacific Conference Series, Vol. 238,
  Astronomical Data Analysis Software and Systems X, ed. F.~R. {Harnden}, Jr.,
  F.~A. {Primini}, \& H.~E. {Payne}, 269

\bibitem[{{Lupton} {et~al.}(2002){Lupton}, {Ivezi\'{c}, \v{Z}.}, {Gunn},
  {Knapp}, {Strauss}, \& {Yasuda}}]{Lupton2002}
{Lupton}, R.~H., {Ivezi\'{c}, \v{Z}.}, {Gunn}, J.~E., {et~al.} 2002, in Society
  of Photo-Optical Instrumentation Engineers (SPIE) Conference Series, Vol.
  4836, Survey and Other Telescope Technologies and Discoveries, ed. J.~A.
  {Tyson} \& S.~{Wolff}, 350--356

\bibitem[{{Martinez} {et~al.}(2010){Martinez}, {Kolb}, {Sarazin}, \&
  {Tokovinin}}]{MartinezMessenger}
{Martinez}, P., {Kolb}, J., {Sarazin}, M., \& {Tokovinin}, A. 2010, The
  Messenger, 141, 5

\bibitem[{{Oya} {et~al.}(2016){Oya}, {Terada}, {Hayano}, {Watanabe}, {Hattori},
  \& {Minowa}}]{subaruSeeing2016}
{Oya}, S., {Terada}, H., {Hayano}, Y., {et~al.} 2016, Experimental Astronomy,
  42, 85, \dodoi{10.1007/s10686-016-9501-6}

\bibitem[{Peterson~et al.(2015)}]{phosim}
Peterson~et al., J.~R. 2015, ApJS, 218, 14

\bibitem[{{Pier} {et~al.}(2003){Pier}, {Munn}, {Hindsley}, {Hennessy}, {Kent},
  {Lupton}, \& {Ivezi{\'c}}}]{Pier2003}
{Pier}, J.~R., {Munn}, J.~A., {Hindsley}, R.~B., {et~al.} 2003, \aj, 125, 1559,
  \dodoi{10.1086/346138}

\bibitem[{{Racine}(1996)}]{Racine1996}
{Racine}, R. 1996, \pasp, 108, 372, \dodoi{10.1086/133732}

\bibitem[{{Racine}(2009)}]{Racine2009}
{Racine}, R. 2009, in Optical Turbulence: Astronomy Meets Meteorology, ed.
  E.~{Masciadri} \& M.~{Sarazin}, 13--22

\bibitem[{{Roddier}(1981)}]{Roddier1981}
{Roddier}, F. 1981, Progress in optics.~Volume 19.~Amsterdam, North-Holland
  Publishing Co., 1981, p.~281-376., 19, 281,
  \dodoi{10.1016/S0079-6638(08)70204-X}

\bibitem[{{Sesar} {et~al.}(2007){Sesar}, {Ivezi{\'c}}, {Lupton}, {Juri{\'c}},
  {Gunn}, {Knapp}, {DeLee}, {Smith}, {Miknaitis}, {Lin}, {Tucker}, {Doi},
  {Tanaka}, {Fukugita}, {Holtzman}, {Kent}, {Yanny}, {Schlegel}, {Finkbeiner},
  {Padmanabhan}, {Rockosi}, {Bond}, {Lee}, {Stoughton}, {Jester}, {Harris},
  {Harding}, {Brinkmann}, {Schneider}, {York}, {Richmond}, \& {Vanden
  Berk}}]{Sesar2007}
{Sesar}, B., {Ivezi{\'c}}, {\v Z}., {Lupton}, R.~H., {et~al.} 2007, \aj, 134,
  2236, \dodoi{10.1086/521819}

\bibitem[{{Snyder} {et~al.}(2016){Snyder}, {Srinath}, {Macintosh}, \&
  {Roodman}}]{2016SPIE.9906E..42S}
{Snyder}, A., {Srinath}, S., {Macintosh}, B., \& {Roodman}, A. 2016, in
  \procspie, Vol. 9906, Ground-based and Airborne Telescopes VI, 990642

\bibitem[{{Stoughton} {et~al.}(2002){Stoughton}, {Lupton}, {Bernardi},
  {Blanton}, {Burles}, {Castander}, {Connolly}, {Eisenstein}, {Frieman},
  {Hennessy}, {Hindsley}, {Ivezi{\'c}}, {Kent}, {Kunszt}, {Lee}, {Meiksin},
  {Munn}, {Newberg}, {Nichol}, {Nicinski}, {Pier}, {Richards}, {Richmond},
  {Schlegel}, {Smith}, {Strauss}, {SubbaRao}, {Szalay}, {Thakar}, {Tucker},
  {Vanden Berk}, {Yanny}, {Adelman}, {Anderson}, {Anderson}, {Annis},
  {Bahcall}, {Bakken}, {Bartelmann}, {Bastian}, {Bauer}, {Berman},
  {B{\"o}hringer}, {Boroski}, {Bracker}, {Briegel}, {Briggs}, {Brinkmann},
  {Brunner}, {Carey}, {Carr}, {Chen}, {Christian}, {Colestock}, {Crocker},
  {Csabai}, {Czarapata}, {Dalcanton}, {Davidsen}, {Davis}, {Dehnen},
  {Dodelson}, {Doi}, {Dombeck}, {Donahue}, {Ellman}, {Elms}, {Evans}, {Eyer},
  {Fan}, {Federwitz}, {Friedman}, {Fukugita}, {Gal}, {Gillespie}, {Glazebrook},
  {Gray}, {Grebel}, {Greenawalt}, {Greene}, {Gunn}, {de Haas}, {Haiman},
  {Haldeman}, {Hall}, {Hamabe}, {Hansen}, {Harris}, {Harris}, {Harvanek},
  {Hawley}, {Hayes}, {Heckman}, {Helmi}, {Henden}, {Hogan}, {Hogg}, {Holmgren},
  {Holtzman}, {Huang}, {Hull}, {Ichikawa}, {Ichikawa}, {Johnston}, {Kauffmann},
  {Kim}, {Kimball}, {Kinney}, {Klaene}, {Kleinman}, {Klypin}, {Knapp},
  {Korienek}, {Krolik}, {Kron}, {Krzesi{\'n}ski}, {Lamb}, {Leger},
  {Limmongkol}, {Lindenmeyer}, {Long}, {Loomis}, {Loveday}, {MacKinnon},
  {Mannery}, {Mantsch}, {Margon}, {McGehee}, {McKay}, {McLean}, {Menou},
  {Merelli}, {Mo}, {Monet}, {Nakamura}, {Narayanan}, {Nash}, {Neilsen},
  {Newman}, {Nitta}, {Odenkirchen}, {Okada}, {Okamura}, {Ostriker}, {Owen},
  {Pauls}, {Peoples}, {Peterson}, {Petravick}, {Pope}, {Pordes}, {Postman},
  {Prosapio}, {Quinn}, {Rechenmacher}, {Rivetta}, {Rix}, {Rockosi}, {Rosner},
  {Ruthmansdorfer}, {Sandford}, {Schneider}, {Scranton}, {Sekiguchi}, {Sergey},
  {Sheth}, {Shimasaku}, {Smee}, {Snedden}, {Stebbins}, {Stubbs}, {Szapudi},
  {Szkody}, {Szokoly}, {Tabachnik}, {Tsvetanov}, {Uomoto}, {Vogeley}, {Voges},
  {Waddell}, {Walterbos}, {Wang}, {Watanabe}, {Weinberg}, {White}, {White},
  {Wilhite}, {Wolfe}, {Yasuda}, {York}, {Zehavi}, \& {Zheng}}]{SDSSEDR}
{Stoughton}, C., {Lupton}, R.~H., {Bernardi}, M., {et~al.} 2002, \aj, 123, 485,
  \dodoi{10.1086/324741}

\bibitem[{{Tokovinin}(2002)}]{Tokovinin2002}
{Tokovinin}, A. 2002, \pasp, 114, 1156, \dodoi{10.1086/342683}

\bibitem[{{York} {et~al.}(2000){York}, {Adelman}, {Anderson}, {Anderson},
  {Annis}, {Bahcall}, {Bakken}, {Barkhouser}, {Bastian}, {Berman}, {Boroski},
  {Bracker}, {Briegel}, {Briggs}, {Brinkmann}, {Brunner}, {Burles}, {Carey},
  {Carr}, {Castander}, {Chen}, {Colestock}, {Connolly}, {Crocker}, {Csabai},
  {Czarapata}, {Davis}, {Doi}, {Dombeck}, {Eisenstein}, {Ellman}, {Elms},
  {Evans}, {Fan}, {Federwitz}, {Fiscelli}, {Friedman}, {Frieman}, {Fukugita},
  {Gillespie}, {Gunn}, {Gurbani}, {de Haas}, {Haldeman}, {Harris}, {Hayes},
  {Heckman}, {Hennessy}, {Hindsley}, {Holm}, {Holmgren}, {Huang}, {Hull},
  {Husby}, {Ichikawa}, {Ichikawa}, {Ivezi{\'c}}, {Kent}, {Kim}, {Kinney},
  {Klaene}, {Kleinman}, {Kleinman}, {Knapp}, {Korienek}, {Kron}, {Kunszt},
  {Lamb}, {Lee}, {Leger}, {Limmongkol}, {Lindenmeyer}, {Long}, {Loomis},
  {Loveday}, {Lucinio}, {Lupton}, {MacKinnon}, {Mannery}, {Mantsch}, {Margon},
  {McGehee}, {McKay}, {Meiksin}, {Merelli}, {Monet}, {Munn}, {Narayanan},
  {Nash}, {Neilsen}, {Neswold}, {Newberg}, {Nichol}, {Nicinski}, {Nonino},
  {Okada}, {Okamura}, {Ostriker}, {Owen}, {Pauls}, {Peoples}, {Peterson},
  {Petravick}, {Pier}, {Pope}, {Pordes}, {Prosapio}, {Rechenmacher}, {Quinn},
  {Richards}, {Richmond}, {Rivetta}, {Rockosi}, {Ruthmansdorfer}, {Sandford},
  {Schlegel}, {Schneider}, {Sekiguchi}, {Sergey}, {Shimasaku}, {Siegmund},
  {Smee}, {Smith}, {Snedden}, {Stone}, {Stoughton}, {Strauss}, {Stubbs},
  {SubbaRao}, {Szalay}, {Szapudi}, {Szokoly}, {Thakar}, {Tremonti}, {Tucker},
  {Uomoto}, {Vanden Berk}, {Vogeley}, {Waddell}, {Wang}, {Watanabe},
  {Weinberg}, {Yanny}, {Yasuda}, \& {SDSS Collaboration}}]{York2000}
{York}, D.~G., {Adelman}, J., {Anderson}, Jr., J.~E., {et~al.} 2000, \aj, 120,
  1579, \dodoi{10.1086/301513}

\bibitem[{{Ziad} {et~al.}(2000){Ziad}, {Conan}, {Tokovinin}, {Martin}, \&
  {Borgnino}}]{vk2}
{Ziad}, A., {Conan}, R., {Tokovinin}, A., {Martin}, F., \& {Borgnino}, J. 2000,
  \ao, 39, 5415, \dodoi{10.1364/AO.39.005415}

\end{thebibliography}
